\documentclass{PoS}
\usepackage{subfigure}
\usepackage{amssymb}
\usepackage{fontenc}
\usepackage{times}
\usepackage{mathptmx}
\newcommand{\farcs}{\mbox{\ensuremath{.\!\!^{\prime\prime}}}}   
\newcommand{\fsec}{\mbox{\ensuremath{.\!\!^{\mathrm{s}}}}}      
\newcommand{\arcdeg}{\ensuremath{^{\circ}}}                     
\newcommand{\sun}{\ensuremath{_{\odot}}}                        

\newcommand{\araa}{ARA\&A}

\newcommand{\baas}{BAAS}

\title{Three epochs of EVN observations towards IRAS 23365+3604}

\ShortTitle{Three epochs of EVN observations towards IRAS 23365+3604}

\author{\speaker{Cristina Romero-Ca\~nizales}%
         \thanks{We acknowledge support from grant AYA2009-13036-C02-01, sponsored by the Spanish MICINN. M.A.P.T acknowledges
         support from the Autonomic Government of Andalusia under grants P08-TIC-4075 and TIC-126. This research has also benefited from 
         research funding from the  European Community Framework Programme 7, Advanced Radio Astronomy in Europe, grant agreement 
         no.: 227290. Thanks to JIVE and especially to Zsolt Paragi and Bob Campbell  for their invaluable help in this project. \emph{The European 
         VLBI Network is a joint facility of European, Chinese, South African and other radio astronomy institutes funded by their national research 
         councils.}}\\
        Instituto de Astrof\'{\i}sica de Andaluc\'{\i}a - CSIC, 18008 Granada, Spain \\
        E-mail: \email{cromero@iaa.es}}

\author{Miguel \'Angel P\'erez-Torres\\
        Instituto de Astrof\'{\i}sica de Andaluc\'{\i}a - CSIC, 18008 Granada, Spain \\
        E-mail: \email{torres@iaa.es}}

\author{Antxon Alberdi\\
        Instituto de Astrof\'{\i}sica de Andaluc\'{\i}a - CSIC, 18008 Granada, Spain \\
        E-mail: \email{antxon@iaa.es}}

\abstract{The European VLBI Network (EVN) provides us with the necessary sensitivity and angular resolution to study the 
nuclear and circumnuclear regions in Luminous (L$_{\mathrm{FIR}} > 10^{11}$~L$\sun$) and Ultraluminous 
(L$_{\mathrm{FIR}} > 10^{12}$~L$\sun$) Infrared Galaxies. The high Star Formation Rates (SFR) inferred for
these galaxies implies both the presence of a high number of massive stars and a dense surrounding medium.
Therefore, bright radio SNe are expected to occur. With the aim of estimating the SFR in ULIRGs by means of
Core Collapse supernova (CCSN) detections, we started an observing campaign with the EVN on a small sample of 
the brightest and farthest ULIRGs in the local Universe. We present here our results from three epochs of 
quasi-simultaneous observations with the EVN at 6 and 18~cm towards one of the objects in our sample: 
IRAS 23365+3604.}

\FullConference{10th European VLBI Network Symposium and EVN Users Meeting: VLBI and
the new generation of radio arrays \\
                 September 20-24, 2010\\
                 Manchester, UK}

\begin{document}

\section{IRAS 23365+3604}

IRAS 23365+3604 ($\alpha=$23$^h$39$^m$01$\fsec$7, $\delta=$36$\arcdeg21^{\prime}14^{\prime\prime}$),
hereafter IRAS 2336, is an advanced merger 
at a distance of 252~Mpc (1~mas$\approx$0.8~pc) with a very high luminosity 
(log(L$_{\mathrm{FIR}} /$L$_\odot $) $=$ 12.13) which corresponds to a CCSN rate of $\approx5$~SN/year \cite{condon}.

\section{Observations and their analysis}

We present multi-epoch, -frequency EVN observations of the ULIRG IRAS 2336 (see Figure \ref{fig:epochs}). 
Putative young radio SNe with typical peak luminosities of L$\sim$10$^{27-28}$~erg~s$^{-1}$Hz$^{-1}$ should have a flux of  
0.01 to 0.1 mJy at the distance of IRAS 2336. The imaging process and therefore the analysis of the observations has proved to 
be a very challenging task due to:
\begin{itemize}
\item The presence of strong extended emission ($\sim$10~mJy, $\Theta$=0$\farcs$5) which hinders the detection
of compact sources (SNe). 
\item The lack of short baselines, which make the imaging algorithm to fail (compare the images we present here with the images
we presented in the previous EVN Symposium \cite{evn08}).
\end{itemize}
The extended emission at the distance of IRAS 2336 cannot be resolved out with the available angular resolution, 
and the only way to properly map it is by having a good combination of short baselines (which give enough information for
closure phase) and/or using a combination of Gaussian model fitting and delta components. This can be done within the Caltech
imaging programme DIFMAP  \cite{difmap}, which we have used for the imaging process. We then exported the clean images back 
into AIPS to analyse them and to produce the final maps that we present here.

\section{Results}

\begin{figure}
  \begin{center}
    \subfigure[]{
         \label{fig:ir23c1}
         \includegraphics[angle=0, width=4.5cm, height=4.5cm]{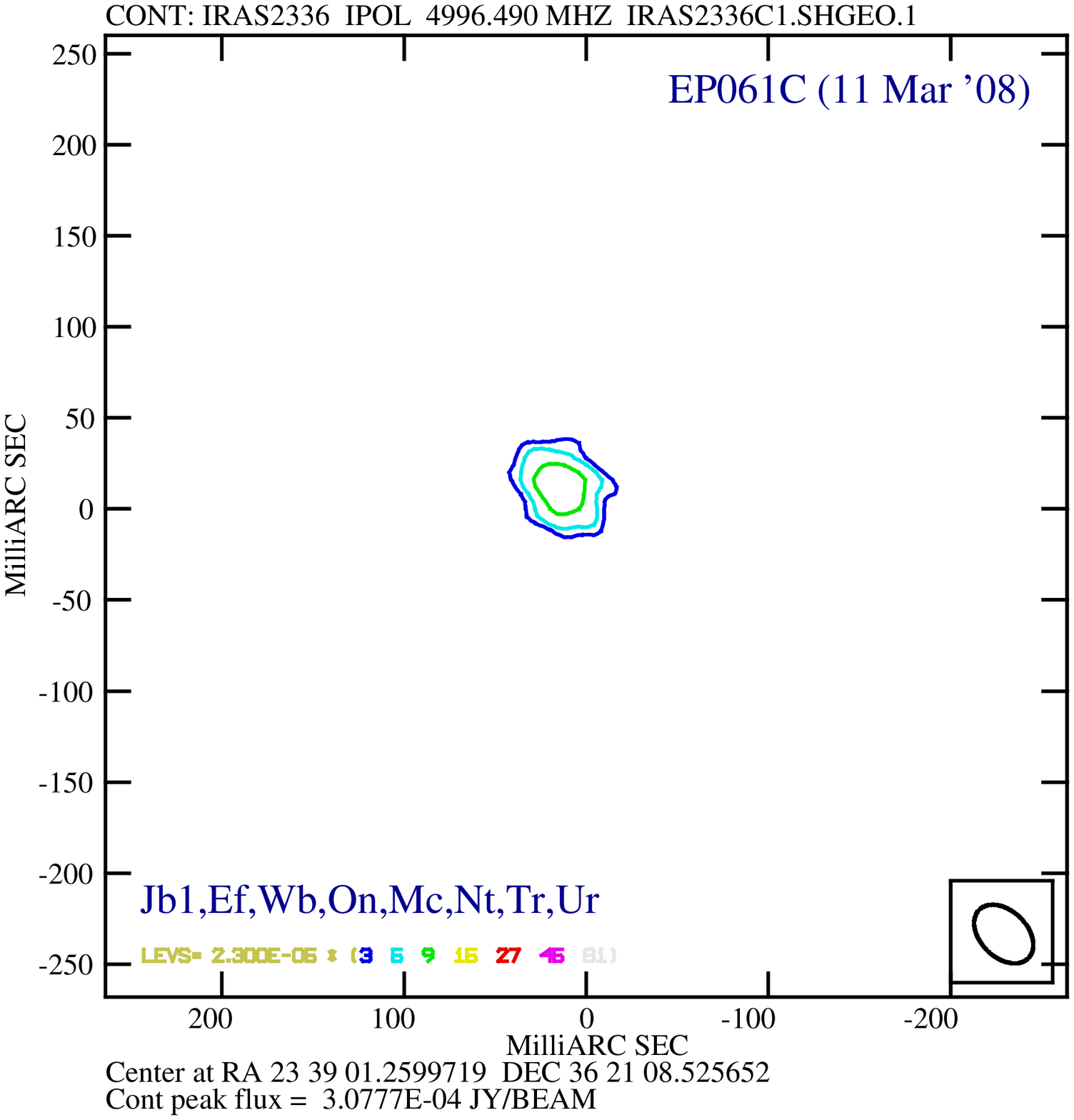}}
    \subfigure[]{
         \label{fig:ir23l1}
         \includegraphics[angle=0, width=4.5cm, height=4.5cm]{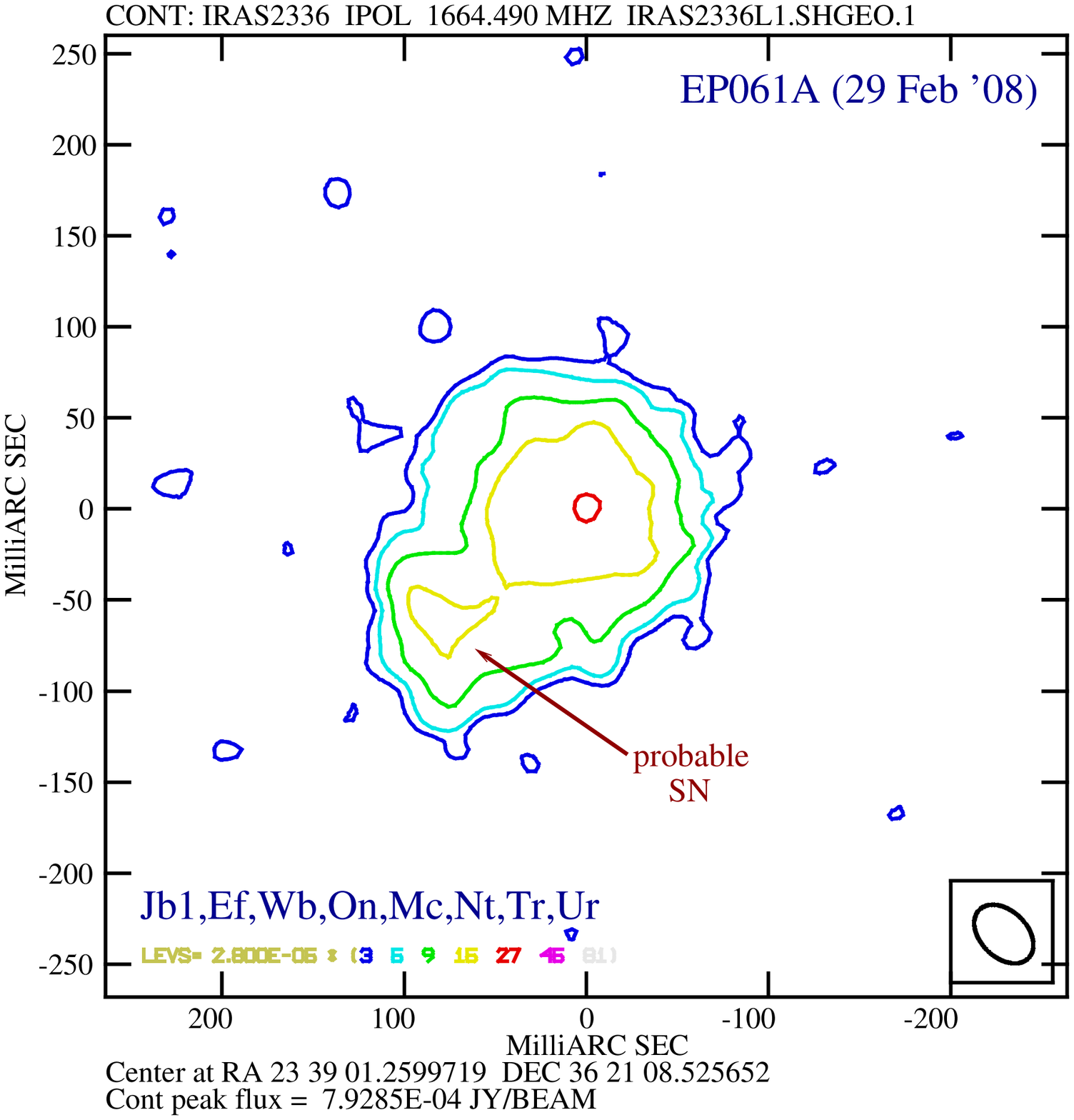}}         
    \subfigure[]{
         \label{fig:ir23cl1}
         \includegraphics[angle=0, width=4.5cm, height=4.5cm]{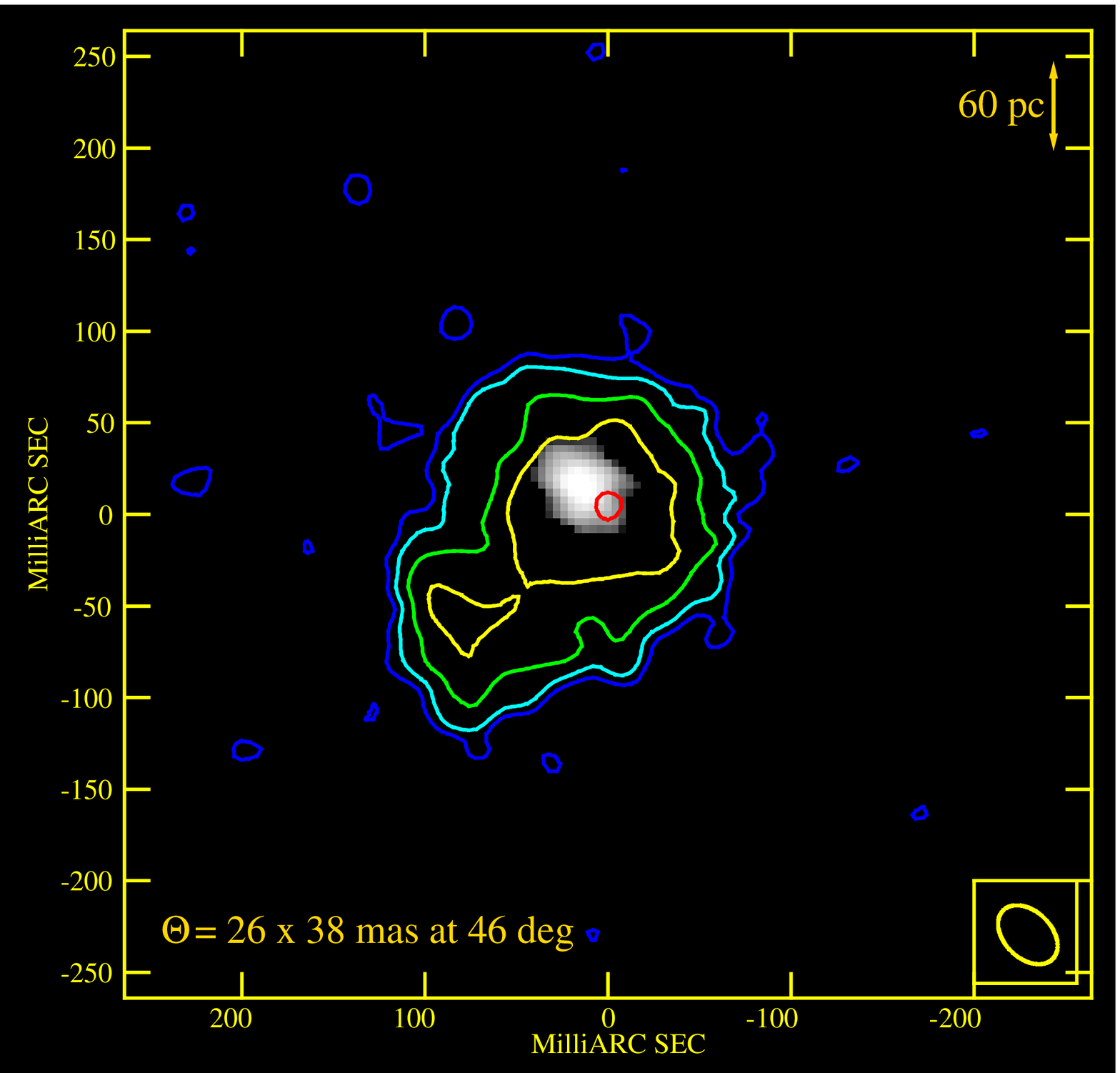}} 
            \vspace{.1in}
       \subfigure[]{
         \label{fig:ir23c2}
         \includegraphics[angle=0, width=4.5cm, height=4.5cm]{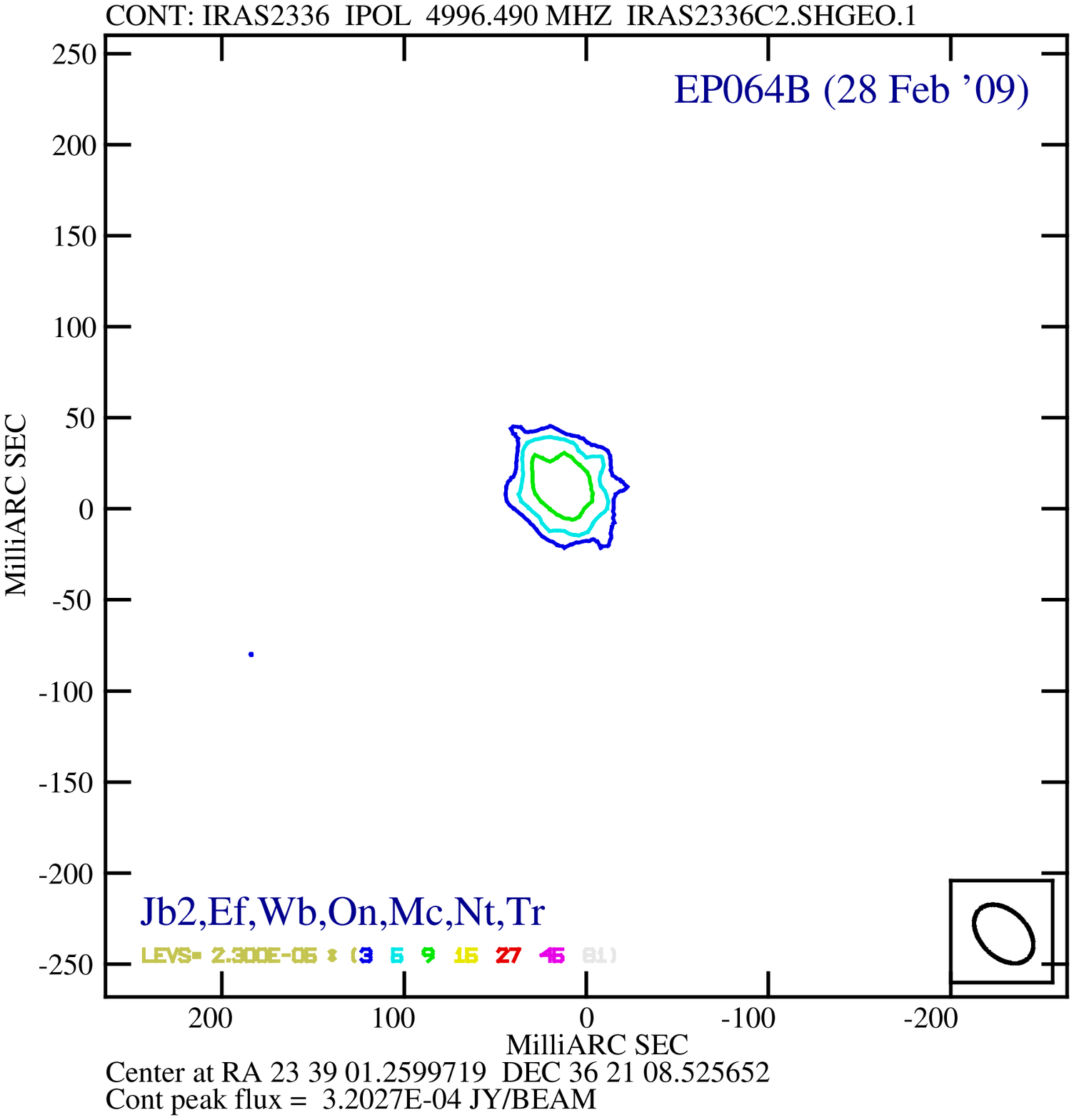}}
    \subfigure[]{
         \label{fig:ir23l2}
         \includegraphics[angle=0, width=4.5cm, height=4.5cm]{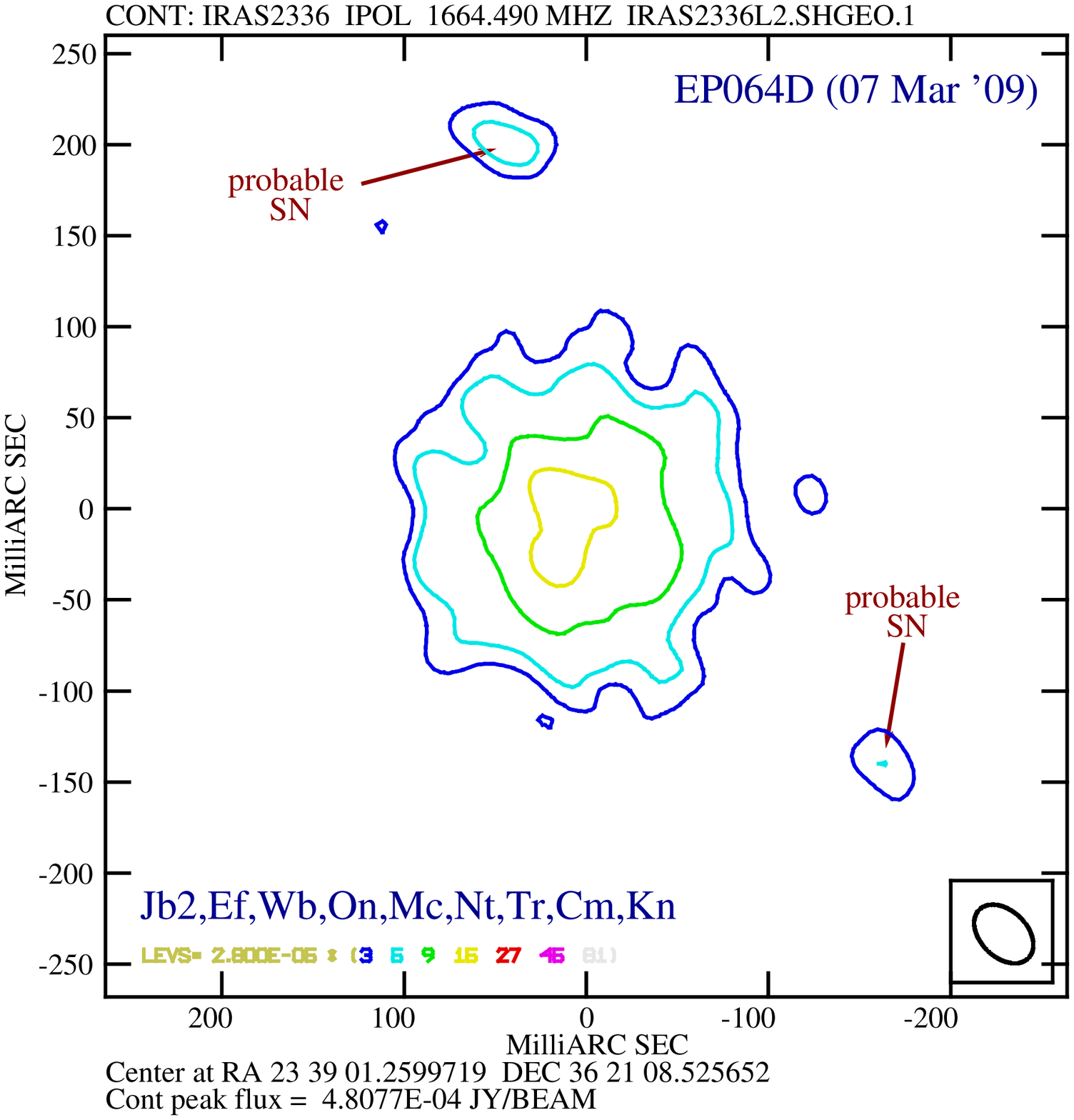}}         
    \subfigure[]{
         \label{fig:ir23cl2}
         \includegraphics[angle=0, width=4.5cm, height=4.5cm]{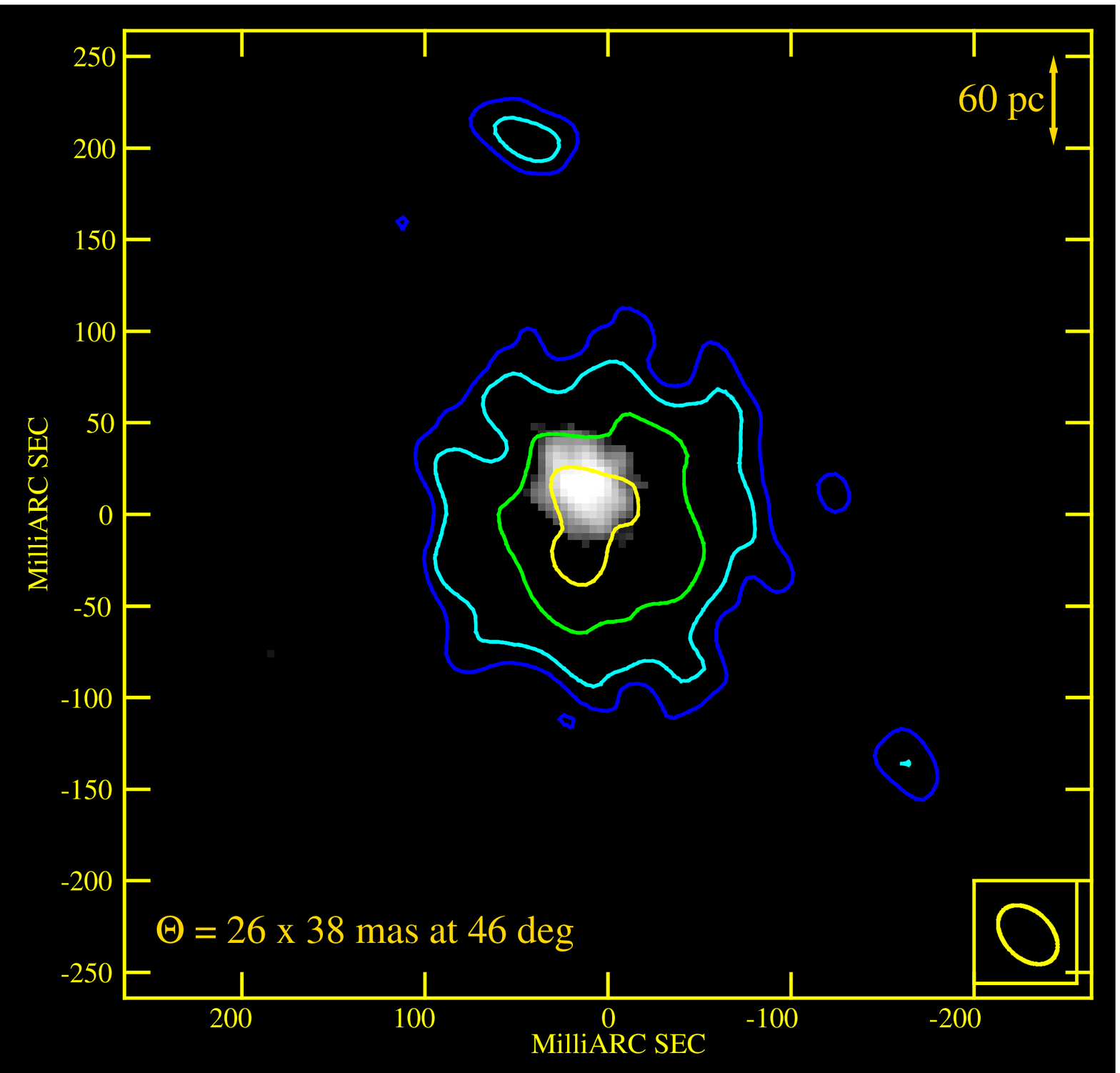}} 
            \vspace{.1in}
    \subfigure[]{
         \label{fig:ir23c3}
         \includegraphics[angle=0, width=4.5cm, height=4.5cm]{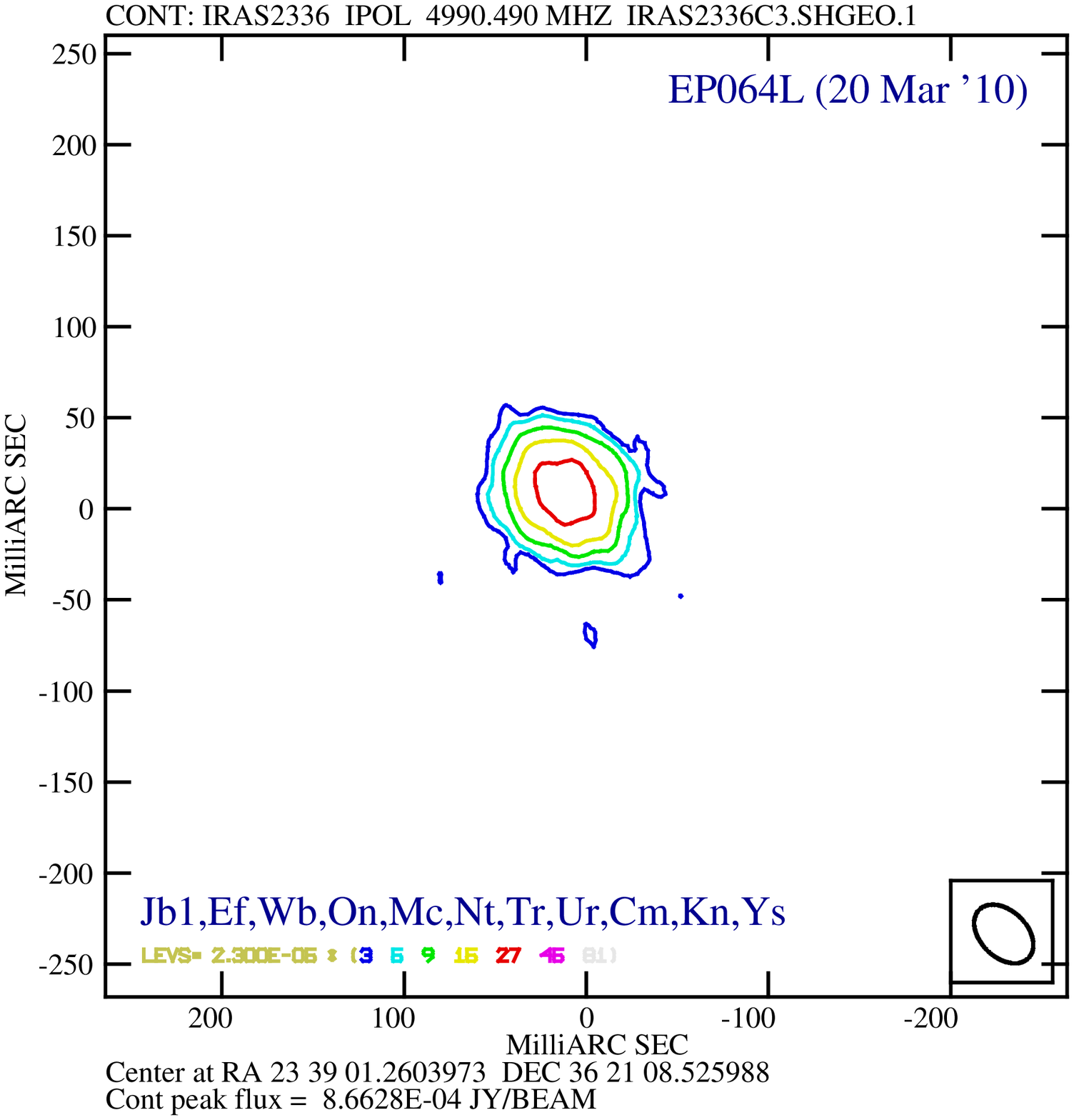}}
    \subfigure[]{
         \label{fig:ir23l3}
         \includegraphics[angle=0, width=4.5cm, height=4.5cm]{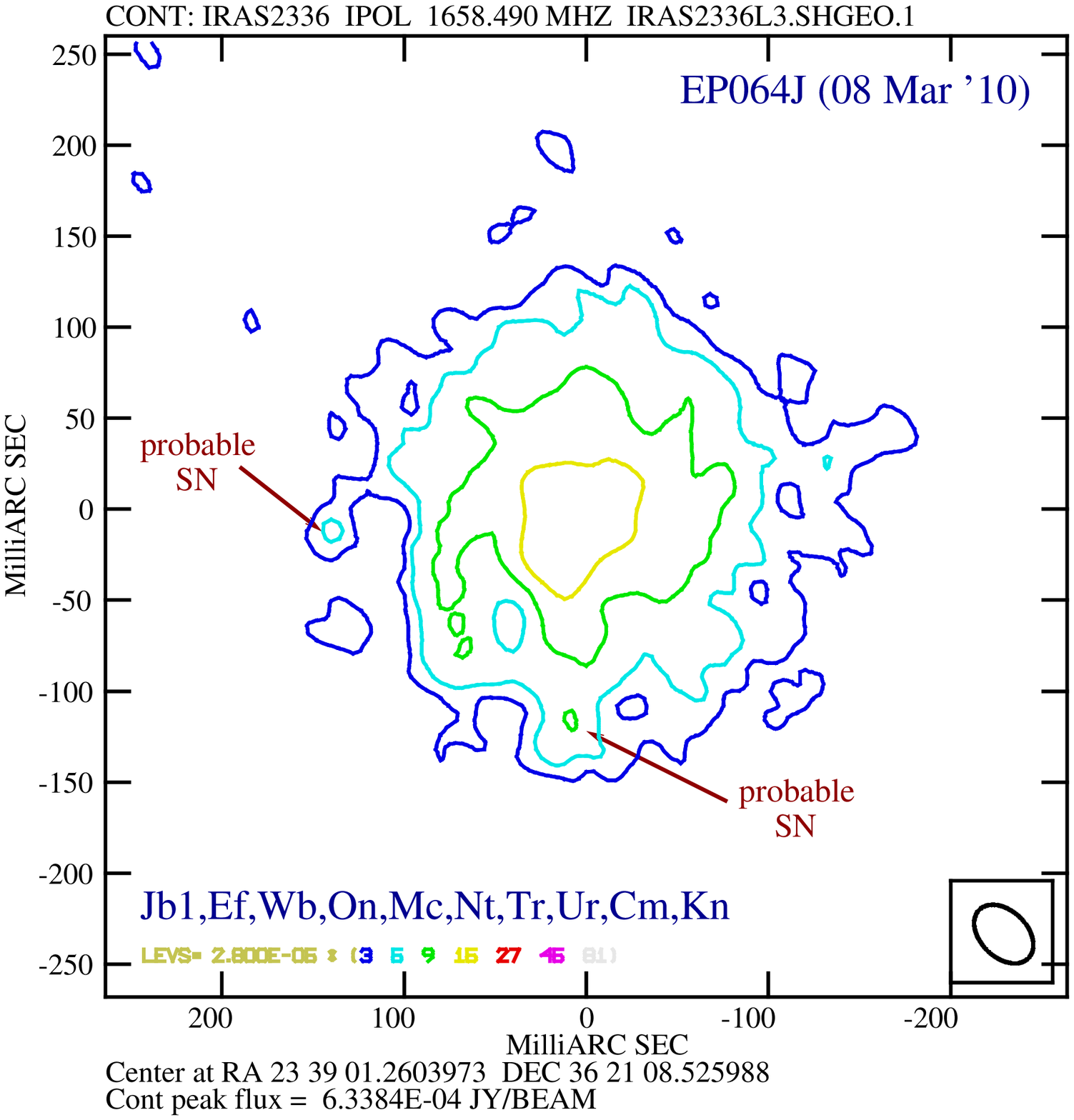}}         
    \subfigure[]{
         \label{fig:ir23cl3}
         \includegraphics[angle=0, width=4.5cm, height=4.5cm]{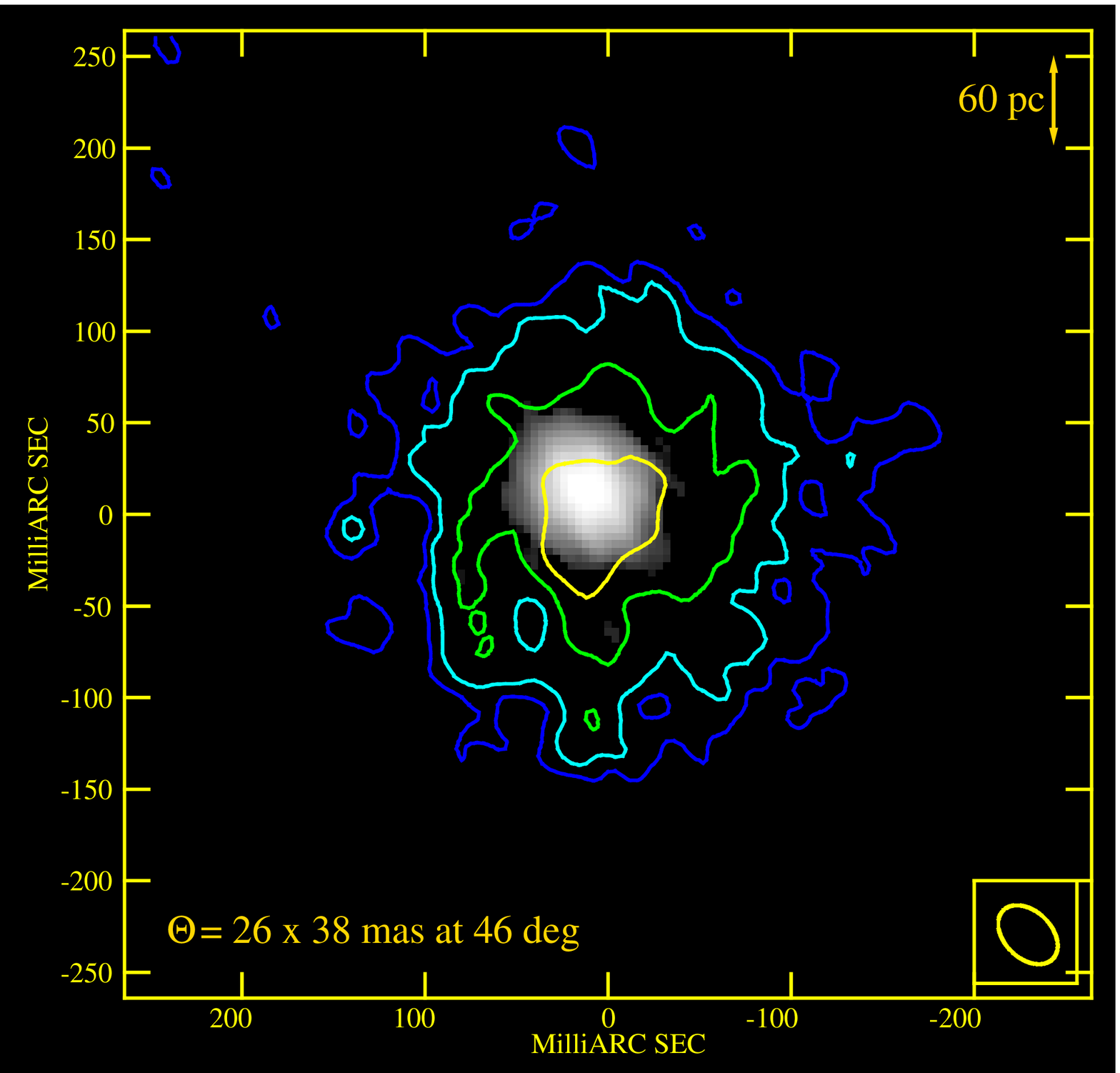}} 
   \caption { \footnotesize{IRAS 2336 at 6~cm (left), 18~cm (middle) and 18~cm contours overlaid on grey scale 6 cm images (right), in three 
   different epochs (top to bottom): March 2008, February 2009 and March 2010. All the images have been degraded to the epoch with lowest 
   resolution (18 cm, second epoch), and thus convolved with a beam size $\sim$30$\times$45 mas. The lowest contours are $\sim$70 and 
   $\sim$85 $\mu$Jy at 6 and 18~cm, respectively. Sources labelled as probable SN in  Figures (b), (e) and (h), are $>5\sigma$ detections  
   and have inferred luminosities $\sim$10$^{27-28}$~erg~s$^{-1}$Hz$^{-1}$. }}
   \label{fig:epochs}
  \end{center}
\end{figure}

\begin{figure}
\begin{center}
\includegraphics[angle=0, width=11.7cm, height=8.3cm]{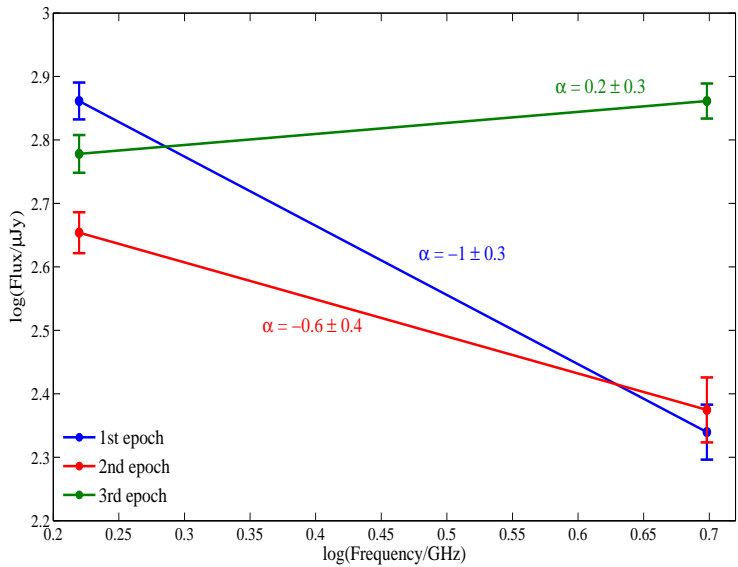}
\caption{\footnotesize{Spectral index ($S\sim\nu^{\alpha}$) evolution of the peak component in the nuclear zone, which we expect 
to be composed by non-thermal sources (e.g. SNe).}}
\label{fig:spec}
\end{center}
\end{figure}

Our images reveal the presence of a nuclear starburst with an approximate extension of 200~pc. We note that at 18~cm, the
size of the nuclear region is larger than at 6~cm, for all the epochs. This can be explained by the longer lifetime of electrons at lower 
frequencies \cite{pachol}. The high luminosities of the nuclear region (see Table \ref{tab:flux}) can only be explained by the 
presence of non-thermal emitters -most likely SNe- within extended emission. In fact, the flattening (from the first to the third epoch) of the spectral index from the 
nuclear zone (see Figure \ref{fig:spec}) can be explained by the variation in flux of sources therein and/or appearance of new sources 
(SN) which would be seen first at higher frequencies and later at lower frequencies \cite{weiler}.
We have also detected some sources  at 18~cm with emission $>$5$\sigma$ (L$\sim$10$^{27-28}$~erg~s$^{-1}$Hz$^{-1}$) with no
counterpart at 6~cm. These are candidates of CCSNe exploding in the circumnuclear region (see sources labelled as probable
SNe in Figure \ref{fig:epochs}).

\begin{table}
\begin{center}
\begin{tabular}{|c|c|c|c|c|}
\hline                        
Epoch & \multicolumn{2}{|c|}{L$_{6~\mathrm{cm}} \times 10^{28}$~erg~s$^{-1}$~Hz$^{-1}$}   & \multicolumn{2}{c|}{L$_{18~\mathrm{cm}} \times 10^{28}$~erg~s$^{-1}$~Hz$^{-1}$}     \\
\cline{2-5}
~  & peak & total  & peak & total   \\
\hline						         										       
1st            & 1.7 $\pm$ 0.2  &  1.7 $\pm$ 0.2  &   5.5 $\pm$ 0.4  &  17.6 $\pm$ 0.4     \\
2nd           & 1.8 $\pm$ 0.2  &  2.4 $\pm$ 0.2  &   3.4 $\pm$ 0.3  &  18.9 $\pm$ 0.3     \\
3rd            & 5.5 $\pm$ 0.4  &  7.2 $\pm$ 0.4  &   4.6 $\pm$ 0.3  &  15.8 $\pm$ 0.3     \\
\hline											                      
\end{tabular}
\end{center}
\caption{\footnotesize{Peak and total luminosities of the nuclear zone at 6~cm and 18~cm in the three different
                                    epochs.  \label{tab:flux}}}
\end{table}

\section{Conclusions} 

The EVN has aided to obtain the deepest and highest resolution radio images ever of one of the most distant ULIRGs in the local Universe. 
High resolution is important, but the information provided by short baselines is also necessary to properly map the morphology of the 
diffuse emission in this kind of sources.

\end{document}